\documentclass[conference,twocolumn,letterpaper]{IEEEtran}

\usepackage{amsmath,amsfonts,amssymb,amsthm}
\usepackage{booktabs}
\usepackage{enumerate}
\usepackage{cite}

\usepackage{graphicx}
\graphicspath{{figures/}}
\usepackage{epstopdf}

\usepackage{txfonts}

\DeclareMathOperator*{\argmax}{arg\,max}

\newcommand{\appropto}{\mathrel{\vcenter{
  \offinterlineskip\halign{\hfil$##$\cr
    \propto\cr\noalign{\kern2pt}\sim\cr\noalign{\kern-2pt}}}}}

\newcommand{\vb}[1]{\mathbf{#1}}

\newcommand{\vbb}{\vb{b}}
\newcommand{\vbc}{\vb{c}}

\newcommand{\vbr}{\vb{r}}

\newcommand{\vbx}{\vb{x}}

\newcommand{\Tra}{\mathrm{T}}

\theoremstyle{remark}

\begin{document}
\bstctlcite{IEEE:BSTcontrol}

\title{Approximate Joint MAP Detection \\ of Co-Channel Signals}
\author{\IEEEauthorblockN{Daniel J. Jakubisin and R. Michael Buehrer}
\IEEEauthorblockA{Mobile and Portable Radio Research Group (MPRG), Wireless@VT, \\
Virginia Tech, Blacksburg, Virginia, USA. E-mail: \{djj,buehrer\}@vt.edu}}

\maketitle

\begin{abstract}
We consider joint detection of co-channel signals---specifically, signals 
which do not possess a natural separability due to, for example, the multiple 
access technique or the use of multiple antennas. Iterative joint detection 
and decoding is a well known approach for utilizing the error correction code 
to improve detection performance. However, the joint maximum a posteriori 
probability (MAP) detector may be prohibitively complex, especially in a 
multipath channel. In this paper, we present an approximation to the joint 
MAP detector motivated by a factor graph model of the received signal. The 
proposed algorithm is designed to approximate the joint MAP detector as closely 
as possible within the computational capability of the receiver.
\end{abstract}

\section{Introduction} \label{sec:introduction}
Detection of a desired signal in the presence of one or more interfering 
signals is a prevalent problem in dense wireless communication systems.
As a result, designing receivers capable of detection in the presence of 
interference has been a very active area of research with numerous algorithms 
proposed in literature. 
Iterative multiuser detection problems have been considered for code 
division multiple access (CDMA) \cite{Wang1999,Boutros2002}, 
spatial multiplexing \cite{Hochwald2003,Haykin2004,Visoz2003}, and 
multiuser MIMO \cite{Lee2012,Hammarberg2012}, among others. 
Receiver algorithms generally fit into one of three categories: linear 
filtering, interference cancellation, and joint detection. 

Linear filtering may be applied in the time, space, or space-time dimension(s) 
and includes techniques such as matched filtering, minimum mean square error 
(MMSE) equalization, and beamforming.
In systems which employ spreading sequences or multiple antennas, linear 
filtering can be an effective means of interference mitigation, specifically 
when the spreading gain or number of antennas is greater than or equal to the 
number of signals present.

Interference cancellation refers to algorithms in which each user's signal is 
canceled from the received signal after detection (e.g., \cite{Arslan2001,Andrews2005} 
and the references therein).
Linear filtering combined with interference cancellation may further improve 
detection and has been a very successful approach for spatial multiplexing 
\cite{Biglieri2007}.
Soft cancellation in conjunction with soft decoding of the channel code---often 
referred to as ``turbo'' interference cancellation---has been shown to 
achieve good results in a CDMA system \cite{Wang1999}. 

Optimal maximum \emph{a posteriori} probability (MAP) detection is performed by 
jointly detecting both the desired and co-channel signals.  
The detection stage is separated from decoding and 
probabilistic information is passed between the joint MAP detector and a 
collection of single user decoders. 
The separation of detection and decoding is justified by 
message passing algorithms which operate on a factor graph of the joint 
probability density function \cite{Wymeersch2007}. 
Yet, even with the separation of detection and decoding, joint MAP 
detection may be prohibitively complex as a result of high-order modulations, 
numerous users, or inter-symbol interference (ISI).  

A challenging case is detection in the presence of non-orthogonal, asynchronous 
interfering signals using a single receive antenna. 
That is, reception of co-channel signals which do not possess a natural 
separability due to a multiple access technique (such as CDMA) or multiple 
antennas.
As a result, linear filtering and interference cancellation are ineffective 
especially when the signal power levels are similar.

Joint MAP detection in such a signal model is developed and studied in \cite{Moon2012}. 
The separability is achieved due to both frame and symbol timing offsets and an 
error correction code. Joint detection which accounts for the strongest ISI 
terms is proposed. Thus, the algorithm is exponentially complex in the number 
of co-channel signals. For this reason its application is limited to 2 users and
BPSK modulation in \cite{Moon2012}.
A large number of users or high-order modulations in addition to ISI due to the 
asynchronous signals makes the optimal joint MAP detector extremely complex.


Jiang and Li consider single antenna interference 
cancellation in a frequency selective, multiple access channel~\cite{Jiang2009}. 
The same channel code, interleaver, and modulation is assumed for all co-channel signals. 
Signal separability is obtained through the independence of each user's 
multipath channel. Jiang and Li propose a concurrent MAP (CMAP) algorithm 
in which a Gaussian approximation is used for 
co-channel interference and MAP equalization for ISI. 
The CMAP algorithm is compared to joint MAP 
detection\footnote{Due to the complexity of the joint MAP detector, this method 
is only evaluated for two users with BPSK modulation in \cite{Jiang2009}.}, the 
Rake Gaussian method proposed in \cite{Ping2003}, and soft interference 
cancellation with MAP equalization. 
While the CMAP algorithm is the state-of-the-art in
addressing the difficult detection problem described above, 
performance is degraded when the Gaussian approximation is made for 
strong co-channel interference terms.

In this paper, we present a new approximation to the joint MAP detector which
is motivated by a factor graph model of the received signal. The 
proposed algorithm is designed to approximate the joint MAP detector as closely 
as possible within the computational capability of the receiver.
The complexity of the algorithm is adjustable and can be set to account for the
capabilities of the receiver, the desired performance, or the difficulty of the 
detection task. 

The paper is organized as follows.  The system model is presented in 
Section \ref{sec:model} followed by development of the MAP detector in 
Section \ref{sec:MAP}.
The complexity of the proposed algorithm is compared with algorithms
from the literature in Section \ref{sec:complexity} and a detailed 
description of the proposed algorithm is provided in Section \ref{sec:algorithm}.
The algorithms are compared via simulation in Section \ref{sec:results} and 
conclusions are drawn in Section~\ref{sec:conclusion}.

\paragraph*{Notation} Let $\vbx$ denote a column vector
$\vbx = [x_0, \ldots, x_{K-1}]^\Tra$. We use the shorthand $\sum_{x_k}$ to denote 
the summation over the domain of $x_k$. Similarly, $\sum_{ \vbx }$ denotes the 
summation over the domain of the vector 
$\vbx$ and $\sum_{ \vbx \backslash x_{k} }$ denotes the 
summation with respect to all variables except $x_k$.

\section{System Model} \label{sec:model}
In this work we consider single antenna reception of $U$ co-channel 
signals (users).
Let the information bits, coded bits, and symbols of the $u$th user be 
denoted by column vectors $\vbb^{(u)}$, $\vbc^{(u)}$, and $\vbx^{(u)}$, 
respectively. We define the collection of these terms for all users as 
\[ \vb{B} = [\vbb^{(1)}, \ldots, \vbb^{(U)}] \]
\[ \vb{C} = [\vbc^{(1)}, \ldots, \vbc^{(U)}] \]
\[ \vb{X} = [\vbx^{(1)}, \ldots, \vbx^{(U)}]. \]
The $n$th sample of the received signal is given by 
\begin{equation}
	r_n = \sum_{u=1}^{U} \sum_{l=0}^{L-1} h_{l}^{(u)} x_{n-l}^{(u)} + w_n,
	\label{eq:signal_model}
\end{equation}
where $\vb{h}^{(u)} = [h_{0}^{(u)}, h_{1}^{(u)}, \ldots , h_{L-1}^{(u)}]^\Tra$ 
denotes the combined effect of the multipath channel and the transmit pulse 
for the $u$th user, $L$ is the number of channel taps, and 
$\{w_n\}_{n=0}^{N-1}$ are independent and identically distributed circularly-symmetric 
complex Gaussian random variables with variance $\sigma^2$. 
The collection of all received samples is denoted 
$\vbr = [r_0, \ldots, r_{N-1}]$.
In general the transmitted signals may be symbol-asynchronous.
For the sake of notational simplicity, the model provided in 
\eqref{eq:signal_model} makes a number of assumptions about the received 
signal---for example, that the channel duration of the users $L$ is identical 
and that the received signal is sampled at a single sample per symbol.  
However, the multiuser detection and equalization
algorithms presented in this paper are applicable to the more general cases.

\section{MAP Detection} \label{sec:MAP}
The goal of the receiver is to detect all information bits $\vb{B}$ given 
observation $\vbr$. Because of the complexity of sequence detection of 
$\vb{B}$, we desire to perform MAP symbol-by-symbol (in our case, bit-by-bit) 
detection. The detector for the $k$th bit of user $u$ is given by
\begin{equation} 
	\hat{b}_{k}^{(u)} = \argmax_{b_{k}^{(u)}} \sum_{ \vb{B} \backslash b_{k}^{(u)} }{p(\vb{B}|\vbr)} ,
	\label{eq:MAP_bit_detection}
\end{equation}
where the marginal is computed for $b_{k}^{(u)}$. According to Bayes' rule, 
\eqref{eq:MAP_bit_detection} is equivalent to 
\begin{equation} 
	\hat{b}_{k}^{(u)} = \argmax_{b_{k}^{(u)}} \sum_{ \vb{B} \backslash b_{k}^{(u)} }{f(\vbr,\vb{B})},
	\label{eq:MAP_bit_bayes}
\end{equation}	
where the term $1/f(\vbr)$ is a constant which has been removed. By the Total 
Probability Theorem, \eqref{eq:MAP_bit_bayes} can further be expressed as a 
marginalization over the full joint distribution as given by
\begin{equation} 
	\hat{b}_{k}^{(u)} = \argmax_{b_{k}^{(u)}} \sum_{ \vb{X}, \vb{C}, \vb{B} \backslash b_{k}^{(u)} }{f(\vbr,\vb{X},\vb{C},\vb{B})}.
	\label{eq:MAP_bit_tpt}
\end{equation}
The marginalization in \eqref{eq:MAP_bit_tpt} cannot be performed directly, 
but an iterative implementation of the sum-product algorithm is well suited 
for this task. 

\subsection{Probability Distribution} \label{sec:distribution}
Taking into account conditional independence of the variables, the joint 
distribution is given by 
\begin{align}
	f(\vbr, \vb{X}, \vb{C}, \vb{B}) = &\prod_{n=0}^{N-1} f(r_n | \vbx^{(1)}, \ldots, \vbx^{(U)}) \notag \\
		&\prod_{u=1}^U{p(\vbx^{(u)} | \vbc^{(u)})p(\vbc^{(u)} | \vbb^{(u)})p(\vbb^{(u)})}.
\end{align}
Factorizations of the modulation $p(\vbx^{(u)} | \vbc^{(u)})$ 
and code $p(\vbc^{(u)} | \vbb^{(u)})$ constraints have been explored in 
the literature (see, for example, \cite{Wymeersch2007,Kschischang2001}).
From \eqref{eq:signal_model} the likelihood function for each term $r_n$ is 
dependent on a subset of the symbols.  We define, 
$\vbx_{[n]}^{(u)} = [x_{n-L+1}^{(u)}, \ldots, x_{n}^{(u)}]^\Tra$ to denote 
the symbols from user $u$ which have components in the $r_n$ sample. 
The distribution is then given by 
\begin{align}
	f(\vbr, \vb{X}, \vb{C}, \vb{B}) = &\prod_{n=0}^{N-1} f(r_n | \vbx_{[n]}^{(1)}, \ldots, \vbx_{[n]}^{(U)}) \notag \\
		&\prod_{u=1}^U{p(\vbx^{(u)} | \vbc^{(u)})p(\vbc^{(u)} | \vbb^{(u)})p(\vbb^{(u)})}.
	\label{eq:joint_dist}
\end{align}

Soft output MAP equalization of an ISI channel may be accomplished via the 
BCJR algorithm \cite{Bahl1974}.  This algorithm was extended to the case of 
joint detection of a desired and co-channel signal in ISI by Moon and Gunther 
\cite{Moon2012}.  The algorithm relies on the introduction of state variables 
$\vb{m}_0,\ldots, \vb{m}_N$ into the likelihood function as follows:
\begin{align}
	&\prod_{n=0}^{N-1} f(r_n | \vbx_{[n]}^{(1)}, \ldots, \vbx_{[n]}^{(U)}) \notag \\
	&\qquad = \prod_{n=0}^{N-1} \sum_{\vb{m}_n} f(r_n, \vb{m}_{n+1} | x_{n}^{(1)}, \ldots, x_{n}^{(U)},\vb{m}_n) p(\vb{m}_0)
	\label{eq:ssm_factor}
\end{align}
where $\vb{m}_n=[ x_{n-L+1}^{(1)}, \ldots, x_{n-1}^{(1)}, \ldots, x_{n-L+1}^{(U)}, \ldots, x_{n-1}^{(N)} ]^\Tra$.

At a high level, local marginals for the symbols are computed by a forward and
backward pass of the BCJR algorithm (also known as the forward-backward 
algorithm). The forward messages are given by
\begin{equation}
	\alpha(\vb{m}_{i+1}) = \prod_{n=0}^{i} \sum_{\vb{m}_n} f(r_n, \vb{m}_{n+1} | x_{n}^{(1)}, \ldots, x_{n}^{(U)},\vb{m}_n) p(\vb{m}_0)
\end{equation}
and
\begin{equation}
	\beta(\vb{m}_i) = \prod_{n=i}^{N-1} \sum_{\vb{m}_{n+1}} f(r_n, \vb{m}_{n+1} | x_{n}^{(1)}, \ldots, x_{n}^{(U)},\vb{m}_n).
\end{equation}
The messages may be defined recursively as given by
\begin{equation}
	\alpha(\vb{m}_{i+1}) = \sum_{\vb{m}_i} \gamma(\vb{m}_{i+1},\vb{m}_i) \alpha(\vb{m}_{i})
\end{equation}
and
\begin{equation}
	\beta(\vb{m}_{i}) = \sum_{\vb{m}_{i+1}} \gamma(\vb{m}_{i+1},\vb{m}_i) \beta(\vb{m}_{i+1}),
\end{equation}
where $\gamma(\vb{m}_{i+1},\vb{m}_i) = f(r_i, \vb{m}_{i+1} | x_{i}^{(1)}, \ldots, x_{i}^{(U)},\vb{m}_i)$, $\alpha(\vb{m}_0) = p(m_0) = 1$, and $\beta(\vb{m}_{N}) = 1$.  A marginal for a particular symbol $x_i^{(u)}$ is given by
$\sum_{x_{[i]}^{(1)}, \ldots, x_{[i]}^{(U)} \backslash x_i^{(u)}} \alpha(\vb{m}_i) \gamma(\vb{m}_{i},\vb{m}_{i+1}) \beta(\vb{m}_{i+1})$.
The joint MAP detector is developed for the case of a received signal with two 
samples per symbol in \cite{Moon2012}.

\begin{figure}[t]
	\centering
	\includegraphics[width=0.8\columnwidth]{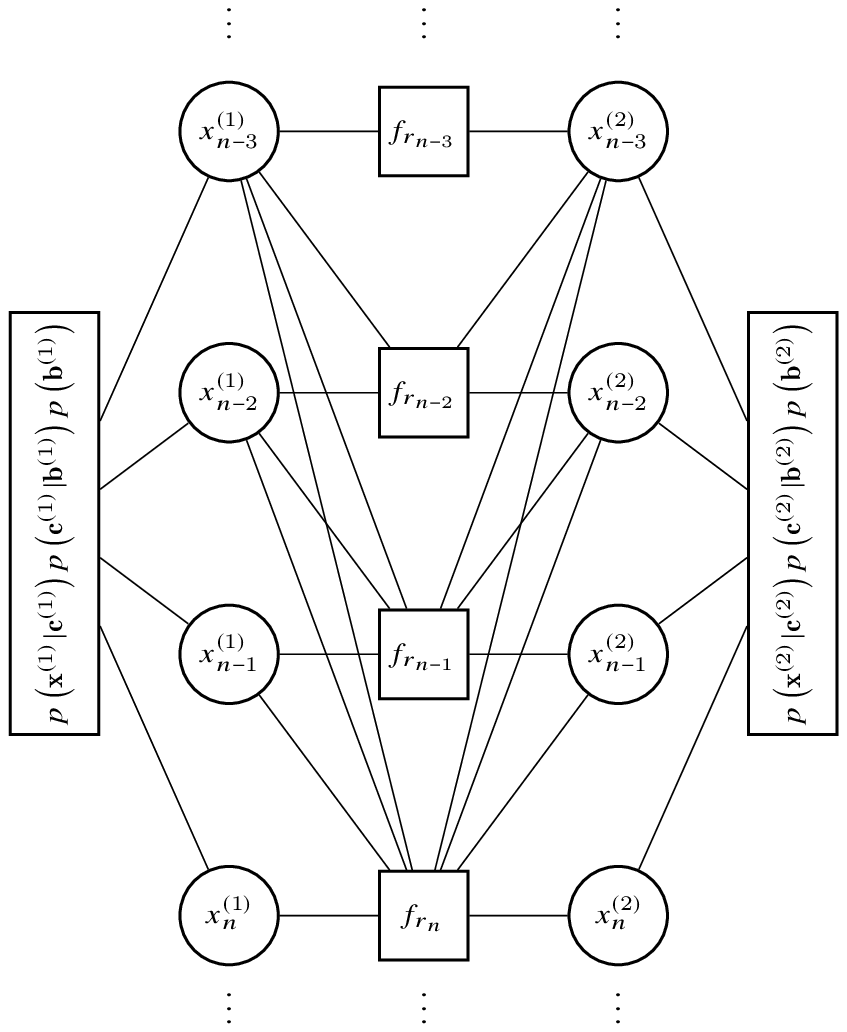}	
	\caption{Factor graph of $f(\vbr,\vb{X},\vb{C},\vb{B})$ for $U=2$ and $L=4$
		based on the factorization in \eqref{eq:joint_dist}.}
	\label{fig:fg_full}
\end{figure}
\begin{figure}[t]
	\centering
	\includegraphics[width=0.8\columnwidth]{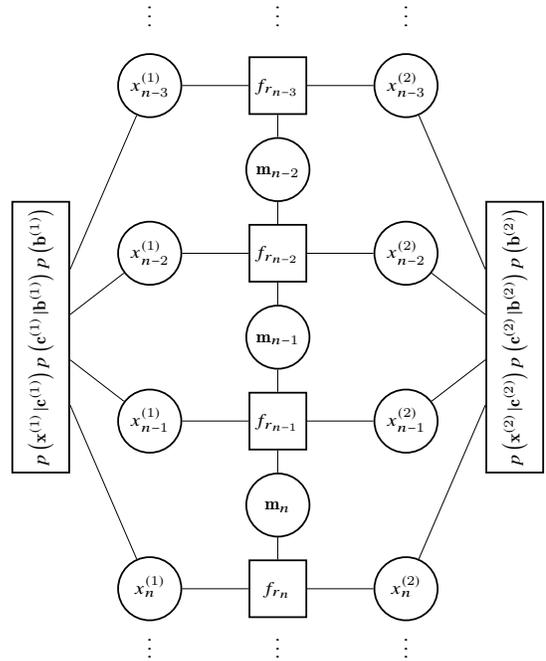}	
	\caption{Factor graph of $f(\vbr,\vb{X},\vb{C},\vb{B})$ for $U=2$ and $L=4$ 
		based on the state space model factorization of \eqref{eq:ssm_factor} 
		substituted into \eqref{eq:joint_dist}.}
	\label{fig:fg_ssm}
\end{figure}

\subsection{Factor Graph Model} \label{sec:graph}
The sum-product algorithm performs efficient marginalization by exploiting 
the factorization of the joint distribution $f(\vbr,\vb{X},\vb{C},\vb{B})$.
As an example, consider the case of $U=2$ and $L=4$.
The factor graph of the joint distribution in \eqref{eq:joint_dist} is given in
Fig.~\ref{fig:fg_full}. 
Similarly, the factor graph of the joint distribution
with the introduction of the state variables is shown in Fig.~\ref{fig:fg_ssm}.
In Fig.~\ref{fig:fg_full}, $f_{r_n}$ denotes the factor 
$f(r_{n}|\vbx_{[n]}^{(1)},\vbx_{[n]}^{(2)})$ and, in Fig.~\ref{fig:fg_ssm}, 
$f_{r_n}$ denotes the factor 
$f(r_n, \vb{m}_{n+1} | x_{n}^{(1)}, \ldots, x_{n}^{(U)},\vb{m}_n)$. 
We refer to the factor graphs in Fig.~\ref{fig:fg_full} and \ref{fig:fg_ssm}
as the \emph{fully connected} graph and the 
\emph{state-space model} (SSM) graph, respectively.

The generalization of the BCJR algorithm to the factor graph of the joint 
distribution is given by the sum-product algorithm \cite{Kschischang2001}. 
The factor nodes 
$p(\vbx^{(u)} | \vbc^{(u)})p(\vbc^{(u)} | \vbb^{(u)})p(\vbb^{(u)})$ 
are further factored when implementing the sum-product algorithm. 
The factor nodes related to the observations $f_{r_n}$ and the symbol
variable nodes make up the ``detection block'' of the factor graph. 
The fully connected graph contains cycles within the detection block; the 
SSM graph eliminates these cycles.  
Cycles have a negative impact on the convergence of the sum-product algorithm.
In Section~\ref{sec:algorithm}, we develop an algorithm
to reduce the complexity of joint MAP detection based on the fully connected 
factor graph of Fig.~\ref{fig:fg_full}.
In Section~\ref{sec:results}, we quantify the loss in performance when 
performing message passing on the fully connected graph versus the SSM graph.

\section{Complexity} \label{sec:complexity}
For both graphs, the complexity associated with \emph{each} of the detection 
factor nodes is $\mathcal{O}(M^{UL})$ where $M$ is the modulation order
of the symbols (assumed to be the same for each user).  
The complexity is exponential in the number of users and channel taps and 
therefore complexity prohibits use of the joint MAP detector in many potential 
co-channel signal scenarios.  
Specifically when either $M>>2$, $U>>2$, or $L>>2$ and especially when this is 
the case for two of these terms.
As an example, the complexity for QPSK, 4 users, and 4 channel taps 
(i.e., $M=4$, $U=4$, and $L=4$) is $\mathcal{O}(10^9)$.

Because of the problem of complexity with joint MAP detector, approaches 
with lower complexity have been considered for this problem.
\begin{itemize}
	\item \emph{Interference Cancellation:} Cancellation may be performed based 
		on either hard or soft decisions. Detection is performed starting with the 
		strongest signal and continuing to the weakest. Soft cancellation may be 
		combined with iterative processing to iteratively improve the soft 
		estimates. 
	\item \emph{Rake Gaussian:} This method was proposed in \cite{Ping2003} for 
		interleave-division multiple access.  In this method, for the detection of 
		symbol $x_{k}^{(u)}$ all other symbols are modeled as Gaussian random 
		variables. This includes the symbols of all other users  
		and all other symbols of the desired user, 
		i.e., $\{x_{k^\prime}^{(u^\prime)}\}_{u^\prime \neq u,k^\prime \neq k}$. 
		The mean and 
		variance of the Gaussian distribution are computed from the extrinsic 
		symbol probabilities obtained from demodulation and decoding. 
	\item \emph{Concurrent MAP (CMAP):} This method was proposed in \cite{Jiang2009} to 
		improve upon the performance of the Rake Gaussian method. In this method, 
		MAP equalization of each user's signal is performed while all other user's 
		signals are modeled as Gaussian random variables. Thus, the complexity of 
		the method is $\mathcal{O}(U\cdot M^L)$, that is, linear in the number of 
		users and exponential in the number of channel taps.
\end{itemize}

\begin{figure}[p]
	\centering
	\includegraphics[width=0.8\columnwidth]{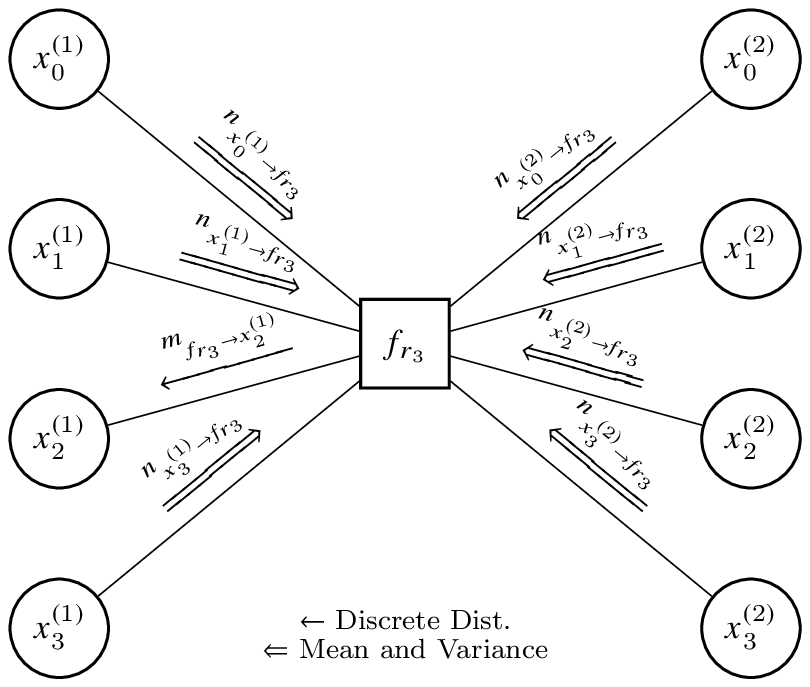}	
	\caption{Factor graph motivated representation of the Rake Gaussian method.} 
	\label{fig:fg_gaussian}
\end{figure}

\begin{figure}[p]
	\centering
	\includegraphics[width=0.8\columnwidth]{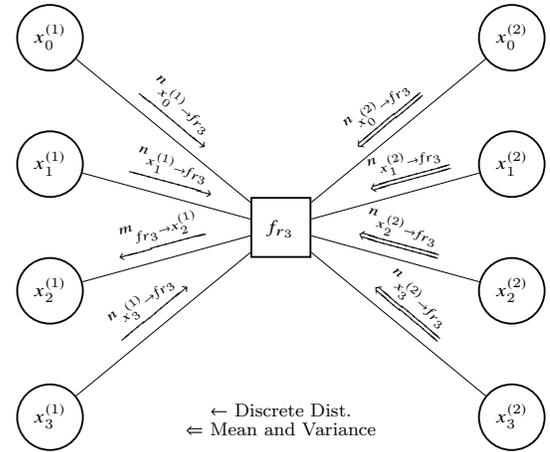}	
	\caption{Factor graph motivated representation of the CMAP method.
		As shown, this factor is a slice of the overall graph to implement MAP 
		equalization of user 1 while modeling the interference from user 2.}
	\label{fig:fg_cmap}
\end{figure}

Visual comparisons of the Rake Gaussian and CMAP algorithms are 
given using factor graphs. The factor node $f_{r_3}$ from the example in 
Fig.~\ref{fig:fg_full} is used to represent the approximations made by the
Rake Gaussian and CMAP algorithms when computing the message
$m_{f_{r_3} \rightarrow x_2^{(1)}}$ in Figs.~\ref{fig:fg_gaussian} and \ref{fig:fg_cmap}, respectively. 
The single arrow represent messages containing discrete
distributions and the double arrow represent messages which contain a 
mean and variance based on a Gaussian approximation.   

\begin{figure}[p]
	\centering
	\includegraphics[width=0.8\columnwidth]{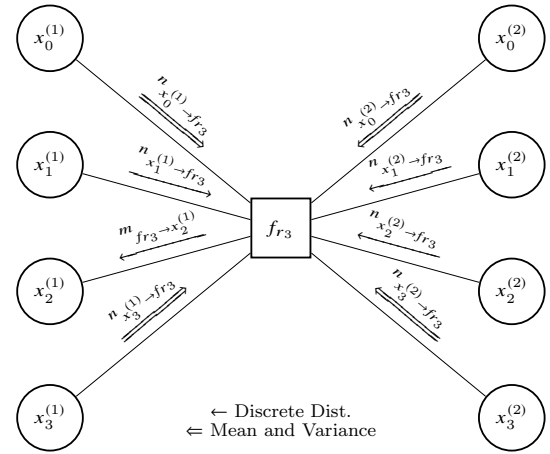}	
	\caption{Factor graph motivated representation of the approximate MAP method 
		developed in this work.}
	\label{fig:fg_hybrid}
\end{figure}

The graphical models of Figs.~\ref{fig:fg_gaussian} and \ref{fig:fg_cmap} 
motivate a new approach in which the distribution of weaker terms in the 
signal component of $r_n$ are modeled as Gaussian random variables. Sum-product message 
passing is performed for the stronger terms in $r_n$. This hybrid approach has 
a complexity determined by the number of messages with discrete distributions
and maintains a single, connected graph. 
The graphical model for the hybrid approach is shown 
in Fig.~\ref{fig:fg_hybrid} where symbols 
$x_{1}^{(1)}$, $x_{2}^{(1)}$, $x_{1}^{(2)}$, and $x_{2}^{(2)}$ are the 
strongest component in $r_3$ for users 1 and 2 (i.e., the power of the 
channel coefficient $|h_l^{(u)}|^2$ is strongest for these terms). This model is 
motivated by common transmit pulse shapes
which contain the majority of their energy within the center of the pulse and 
multipath channels which often exhibit an exponential decay.
A detailed description of the algorithm is provided in the following section.

\section{Approximate MAP Detection Algorithm} \label{sec:algorithm}
Consider a generic interference model (to represent inter-symbol 
interference, co-channel interference, or both) in which $K$ signal 
components $x_1,x_2, \ldots, x_K$ are received with channel coefficients 
$h_1,h_2, \ldots, h_K$, respectively. 
 The received signal is given by 
\[ y = \sum_{k=1}^{K} h_k x_k + w \]
where $y$ represents one sample of a larger sequence of received samples and
the noise $w$ is modeled as a circularly symmetric complex Gaussian 
random variable with variance $\sigma^2$. The factor associated with the 
received sample $y$ is given by
\[ f(y | x_1,\ldots, x_K) = \mathcal{CN} \Biggl( y; \sum_{k=1}^{K} h_k x_k   \, , \, \sigma^2 \Biggr) \]
where the channel coefficients and the noise power $\sigma^2$ are assumed 
to be known.

The message from factor node $f_y$ to variable node $x_k$ is denoted 
$m_{f_y \rightarrow x_k}$.  Similarly, the message from variable node $x_k$
to factor node $f_y$ is denoted $n_{x_k \rightarrow f_y}$. According to the 
sum-product algorithm, the message $m_{f_y \rightarrow x_i}$ is given by 
\begin{equation}
	m_{f_y \rightarrow x_i}(x_i) = \sum_{\vbx \backslash x_i } f(y | x_1,\ldots, x_K ) \prod_{k\neq i} n_{x_k \rightarrow f_y}(x_k).
	\label{eq:spa_comp}
\end{equation}
The proposed algorithm modifies the sum-product algorithm computations
as follows:
\begin{itemize}
	\item The mean and variance of the input messages are computed according to
	\[ \mu_{x_k} = \sum_{x_k} x_k n_{x_k \rightarrow f_y}(x_k) \]
	\[ \sigma_{x_k}^2 = \sum_{x_k} |x_k - \mu_{x_k} |^2 n_{x_k \rightarrow f_y}(x_k)  \]
	for all $k=1,\ldots,K$.
	
	\item For computation of the outgoing message $m_{f_y \rightarrow x_i}(x_i)$, 
	the remaining variables for $k\neq i$ are sorted by their channel 
	coefficient power $|h_k|^2$. 
	Let the set $\mathcal{A}$ index the variables associated with the strongest 
	channel coefficients.  These variables remain a part of 
	the local marginalization as given in \eqref{eq:spa_comp}.  
	The number of variables in the set $\mathcal{A}$ will depend on the 
	acceptable complexity in implementation.
	The indices of the weaker components are included in the set $\mathcal{B}$ 
	and the distributions of these variables are approximated by Gaussian random 
	variables to eliminate the marginalization over these variables.
	Let the variables associated with sets $\mathcal{A}$ and $\mathcal{B}$ 
	be given by $\vbx_{\mathcal{A}}$ and $\vbx_{\mathcal{B}}$, respectively.
	
	\item The message is computed with the following approximate sum-product 
	computation:
	\[ m_{f_y \rightarrow x_i}(x_i) = \sum_{\vbx_{\mathcal{A}}} \tilde{f}( y | x_i, \vbx_{\mathcal{A}} ) \prod_{k \in \mathcal{A}} n_{x_k \rightarrow f_y}(x_k) \]
	where 
\end{itemize}
\begin{align} 
	&\tilde{f}( y | x_i, \vbx_{\mathcal{A}} ) = \notag \\
		&\mathcal{CN} \Biggl( y; h_i x_i + \sum_{k \in \mathcal{A}} h_k x_k + \sum_{l \in \mathcal{B}} h_l \mu_{x_l} \, , \, \sigma^2 + \sum_{l \in \mathcal{B}} |h_l|^2 \sigma_{x_l}^2 \Biggr).
\end{align}

This algorithm is applied to the computation of the sum-product messages at 
each of the detection factors in Fig.~\ref{fig:fg_full}. The complexity of the 
proposed algorithm for each factor is 
$\mathcal{O}(UL\cdot M^{|\mathcal{A}|+1})$ where $|\mathcal{A}|$ 
is the number of symbols included in the set $\mathcal{A}$.  
Thus, by choosing the size of $\mathcal{A}$, the complexity of the algorithm 
may be adjusted to match the computational capability of the receiver and 
performance requirements.

\begin{figure}
	\centering
	\includegraphics[width=0.9\columnwidth]{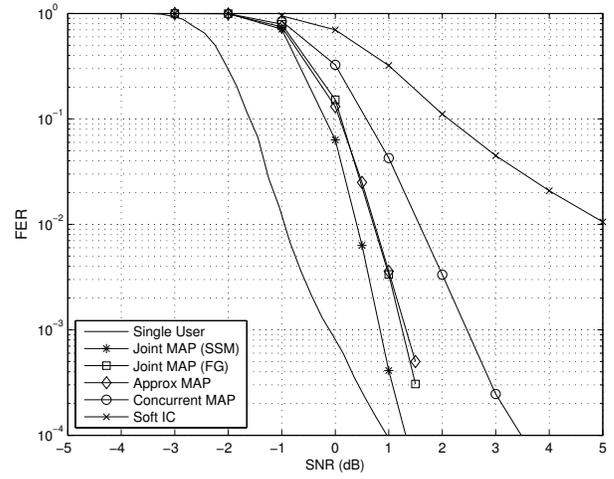}
	\caption{FER comparison of the joint MAP state-space model (Joint MAP (SSM)), 
		the joint MAP fully connected factor graph (Joint MAP (FG)), the proposed 
		approximate MAP algorithm (Approx MAP), CMAP (Concurrent MAP), 
		and soft interference cancellation (Soft IC) algorithms with SIR=0 dB.}
 \label{fig:mud_comp_fer}
\end{figure}
\section{Numerical Results} 	\label{sec:results}
We first simulate the performance for a scenario in line with the one considered in 
\cite{Moon2012}: two users ($U=2$) each employing BPSK modulation ($M=2$). 
ISI results from symbol timing offsets between the users and a transmit pulse
with a duration of four symbol periods ($L=4$). The selection of these 
parameters allows us to simulate the joint MAP detector
for the purpose of comparison.  The simulation parameters are summarized as follows:
\begin{itemize}
	\item Code: 1/2-rate turbo code with 500 coded bits 
	\item Modulation: BPSK
	\item Pulse: Square root raised cosine with $L=4$ and roll-off factor 0.35
	\item A relative time delay between the users of $T/4$ is chosen where $T$ 
		is the symbol period
	\item A relative phase offset between the channel coefficients of the users 
		of $\pi/6$ is chosen
	\item 15 iterations of message passing are performed
\end{itemize}
The FER performance is shown in Fig.~\ref{fig:mud_comp_fer}
for the joint MAP (with the SSM and fully connected factor 
graphs), the proposed novel approach, CMAP, and soft interference 
cancellation algorithms.  
The approximate MAP algorithm is implemented with $|\mathcal{A}|=3$. 
Thus, the proposed approximate MAP algorithm and the CMAP algorithm 
have the same order of complexity (per iteration).
The performance of the fully connected factor graph demonstrates a loss of 
about 0.5 dB compared to the SSM factor graph.
The proposed approximate MAP algorithm is based on the fully connected graph 
and we observe that it achieves nearly identical performance to the receiver 
which uses exact sum-product computations.  
At a FER of $10^{-3}$ the proposed approximate MAP approach and Concurrent 
MAP approach demonstrate losses of 0.5 dB and 1.5 dB, respectively, 
compared to joint MAP detection based on the SSM. We observe that the 
Soft IC method becomes limited by interference as signal-to-noise ratio (SNR) increases.

\begin{figure}
	\centering
	\includegraphics[width=0.9\columnwidth]{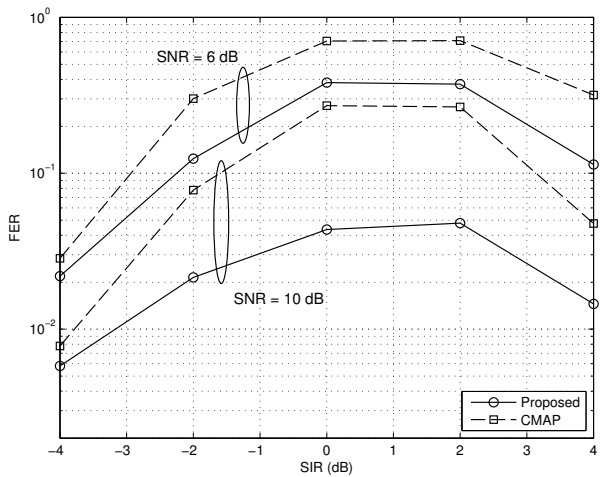}
	\caption{FER of the proposed and CMAP algorithms with respect to SIR. 
	Both signals are detected, and the FER of the desired signal is shown.
	Ten iterations of the receiver are performed.}
 \label{fig:fer_qpsk_sir}
\end{figure}

We also consider a 2-user scenario with QPSK modulation with a 4-tap multipath channel.
The average power in each multipath component is given by 
$[0.644, 0.237, 0.087, 0.032]$.
In Fig.~\ref{fig:fer_qpsk_sir} the FER of the proposed approximate MAP algorithm
and the CMAP algorithm is shown.  
Both SNR and signal-to-interference ratio (SIR) are computed with respect to the instantaneous
power in the multipath channel. 
The most significant improvement in FER is achieved by the proposed approximate MAP 
algorithm when the signals have 
similar power levels ($-3 \leq \text{SIR} \leq 3$ dB) and the SNR is high. 
\begin{figure}
	\centering
	\includegraphics[width=0.9\columnwidth]{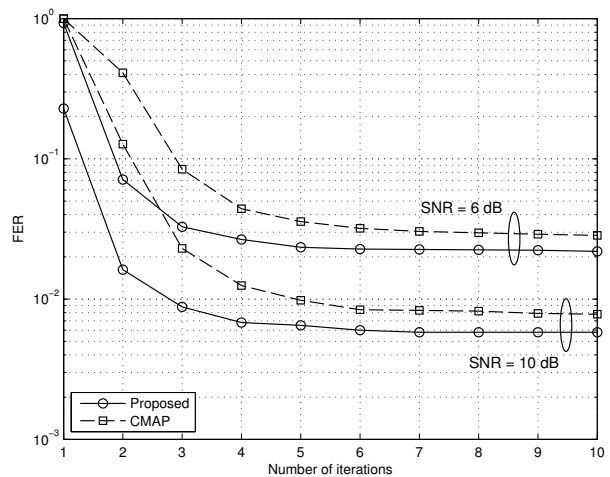}
	\caption{FER of the proposed and CMAP algorithms with respect
	to the number of iterations performed. The SIR is -4 dB.}
 \label{fig:converge_qpsk}
\end{figure}
In Fig.~\ref{fig:converge_qpsk}, the FER is shown with respect to the
number of iterations where we observe that the proposed algorithm 
converges 1-2 iterations faster than CMAP.
Thus, for $|\mathcal{A}|=L-1$, the proposed algorithm reduces 
computational complexity by 20--40\% due to faster convergence.

\section{Conclusion} \label{sec:conclusion}
In this paper, an algorithm is developed which approximates joint MAP 
detection and equalization in co-channel interference.  
The approximate MAP algorithm is based on a fully connected factor graph of 
the joint probability distribution.
The algorithm was shown to operate within 0.5 dB of the joint 
MAP state-space model receiver where the degradation in performance 
was due to the associated factor graph model.
Additionally, the proposed algorithm both improves performance and 
reduces complexity when compared to the state-of-the-art.

\bibliographystyle{IEEEtran}
\bibliography{IEEEabrv,master}

\end{document}